\documentclass[aps,prb,twocolumn,10pt,superscriptaddress,noeprint,longbibliography]{revtex4-1}
\usepackage{amssymb}
\usepackage{amsmath}
\usepackage{graphicx}
\usepackage{dcolumn}
\usepackage{float}
\usepackage{color}
\usepackage{bm}
\usepackage[hidelinks]{hyperref}
\hypersetup{
    colorlinks,
    citecolor=blue,
    filecolor=blue,
    linkcolor=blue,
    urlcolor=blue
}

\newcommand{\hh}{\hat{H}}

\newcommand{\dpsi}{\psi^\dagger}

\newcommand{\im}{\mathrm{i}}
\newcommand{\e}{\textrm{e}}

\newcommand{\dD}{{\rm d}}

\newcommand{\dw}{\delta_{\rm W}}
\newcommand{\dwc}{\delta_{\rm c}}

\usepackage{stackengine}

\begin{document}

\title{Local spin transfer torque and magnetoresistance in domain walls with variable width}

\author{Hamidreza Kazemi}
    \email{hrkazemi.10@gmail.com}
    \affiliation{Physics Department and Research Center OPTIMAS, Technische Universit\"at Kaiserslautern, 67663 Kaiserslautern, Germany}
\author{Sebastian Eggert}
    \affiliation{Physics Department and Research Center OPTIMAS, Technische Universit\"at Kaiserslautern, 67663 Kaiserslautern, Germany}
\author{Nicholas Sedlmayr}
    \email{sedlmayr@umcs.pl}
    \affiliation{Institute of Physics, M.~Curie-Sk{\l}odowska University, 20-031 Lublin, Poland}

\date{\today}

\begin{abstract}
    Use of a spin polarized current for the manipulation of magnetic domain walls in ferromagnetic nanowires has been the subject of intensive research for many years. Recently, due to technological advancements, creating nano-contacts with special characteristics is becoming more and more prevalent. We now present a full quantum investigation of the magnetoresistance and the spin transfer torque in a domain wall, which is embedded in a nano-contact of Ni$_{80}$Fe$_{20}$, where the size of the domain wall becomes a relevant tunable parameter.  The dependence on the domain wall width as well as the spatial dependence of the torque along the domain wall can be analyzed in complete detail.  The magnetoresistance drops with increasing domain wall width as expected, but also shows characteristic modulations and points of resonant spin-flip transmission. The spin transfer torque has both significant in-plane and out-of-plane contributions even without considering relaxation. A closer inspection identifies contributions from the misalignment of the spin density for short domain walls as well as an effective gauge field for longer domain walls, both of which oscillate along the domain wall. 
\end{abstract}

\maketitle

\section{Introduction}

Spintronics has come a long way since Berger\cite{Berger1996} and Slonczewski\cite{Slonczewski1996} independently predicted current induced magnetization dynamics almost two decades ago. The first generation of spin transfer torque (STT) magnetic random access memories (MRAMs) are already commercially available and proposals for other STT-based devices are regularly being put forward.\cite{Manchon2019} Contrary to traditional MRAMs where one uses an Oersted field to manipulate the magnetization, in STT MRAMs the external magnetic field is substituted by a current induced switching mechanism. Current-controlled magnetic domain-wall movement in a nanowire was first shown in Ref.~[\onlinecite{Hayashi2008}]. This effect is also responsible for a host of other phenomena such as spin-wave excitations\cite{Balashov2008,Tserkovnyak2008,Ji2003} and current-driven ferromagnetic resonance.\cite{Kiselev2003,Tulapurkar2005,Kasai2006,Sankey2006,Fuchs2007,Wang2013} The shortcomings in STT-MRAMs are endurance issues due to the out-of-plane writing geometry, and the need for large current densities.\cite{KhaliliAmiri2018}
        
One promising route for improved STT efficiency is miniaturization in order to create a larger spin-misalignment and also harness quantum effects such as interference and resonant tunneling.\cite{Razavy2003}  Great progress has been made in the fabrication of nano-contacts, which can be manufactured down to $\sim1$ nm width.\cite{Reeve2014} Notches in ferromagnetic films can be used to trap a domain wall (DW), where the length and width can be manipulated. DW movement has been observed in these constructions\cite{Heyne2010} and their resistivity has been investigated.\cite{Reeve2019}
    
An approximate theoretical understanding of STT and magnetoresistance is possible using a number of theoretical approaches,\cite{Tserkovnyak2006,Kohno2006,Piechon2007,Tatara2008,Duine2009,Duine2009,Garate2009,Cheng2013,Waintal2003,Tatara2008,Xiao2006,Vanhaverbeke2007,Taniguchi2009} 
which typically have to make reasonable assumptions for the relaxation processes to capture the non-adiabatic contributions in particular materials. In this paper we now put emphasis on the 
quantum mechanical time-evolution over short distances to capture interference effects and oscillations in the local torque, which have so far not received much attention but show interesting resonant tunneling effects and give deeper insight into the microscopic motion.  
Our calculations provide predictions for experimental signatures of corresponding quantum oscillations in nano-contacts and pave the way for further many-body simulations in quasi one-dimensional setups, where remarkable correlation effects have been predicted.\cite{Araujo2006,Araujo2007,Pereira2004,Sedlmayr2011} For long DWs and for special resonant DW widths we find that the spin polarization of the incoming electron is transferred coherently to the DW, i.e.~without any relaxation or production of heat.
As we will show here, scattering of electrons from sharp inhomogeneous magnetic structures creates a spin density which is no longer aligned with the local magnetization direction and leads to significant in-plane and out-of-plane torques. Previous studies have shown that scattering from successive sharp DWs can induce non-aligned spin densities which lower the threshold for current induced DW motion and lead to ordering of the DWs.\cite{Sedlmayr2009,Sedlmayr2011c,Sedlmayr2011a,Golovatski2018}

This paper is organized as follows: In Sec.~\ref{sec:nanocontacts} we present the model and theory to find the STTs in magnetic DWs constrained in nano-contacts, which is used in Sec.~\ref{sec:resistance} to calculate the magnetoresistance as a function of DW width. In Sec.~\ref{sec:spindensity} we calculate the spin density caused by scattering of conduction electrons from the DW, which results in a characteristic behavior of STTs as a function of DW-width and momenta of the incident particle as presented in Sec.~\ref{sec:STT}.  Finally, in Sec.~\ref{sec:conclusions} we conclude.

\section{Domain walls in nanocontacts}\label{sec:nanocontacts}

Nanocontacts in ferromagnetic strips are energetically preferential locations for DWs to locate. The transport across the DW only involves the one-dimensional momentum perpendicular to the wall, while the momentum components parallel to the wall are conserved or simply correspond to a confined standing wave due to the constriction. Therefore, the theoretical description of the scattering is inherently one-dimensional even though the experimental setup may be much more complicated.  While the parallel momentum components are irrelevant in the scattering, they nonetheless must be considered to obtain the correct total energy of the incoming wave, which is typically close to the Fermi energy. This can be taken into account by averaging over all directions as discussed in the next section.
    
We are interested in a wire with bulk ferromagnetic order and a single DW located at the nanocontact. We use the standard $s$-$d$-approximation of conduction electrons interacting with a bulk classical magnetization, which can be regarded as static on the typical electronic time-scales. The $s$-$d$-model has recently been shown to describe Permalloy (Ni$_{80}$Fe$_{20}$) near the Fermi-energy well.\cite{Reeve2019} This is then used to calculate the scattering of the electrons from the non-collinear magnetization. The Hamiltonian $H=H_0+H_M$ on the lattice along the direction of changing magnetization is given in terms of electron creation operators $\dpsi_{j \sigma}$ with spin $\sigma=\pm 1$
    \begin{equation}\label{fullhamiltonian}
    H_0=-\sum_{j \sigma}\left[t\left(\dpsi_{j \sigma}\psi_{j+1 \sigma}+\dpsi_{j+1 \sigma}\psi_{j \sigma}\right)+\mu\dpsi_{j \sigma}\psi_{j \sigma}\right]\,,
    \end{equation}
and an $s$-$d$ coupling to the magnetization
    \begin{equation}\label{fullhamiltonianm}
    H_M=-\frac{J}{2}\sum_{j\sigma \sigma'}\,\dpsi_{j \sigma}\psi_{j \sigma'}\vec{\sigma}_{\sigma\sigma'} \cdot \hat{n}_j\,.
    \end{equation}
Here, $\mu$ is the chemical potential, $t$ is the hopping integral, and $J$ is the $s$-$d$ coupling strength, which are known for Ni$_{80}$Fe$_{20}$ from Ref.~[\onlinecite{Reeve2019}]. We will use $\hbar=1$ throughout. For a N\'eel DW the magnetization is assumed to change along the lattice in the direction of the unit vector $\hat{n}_j$, which can be parameterized by an angle $\Theta_j$ in the $x$-$z$-plane going from up ($\Theta_{-\infty}=0$) to down ($\Theta_{\infty}=\pi$),
    \begin{equation}
    \hat{n}_j=\sin[\Theta_j]\hat{\mathbf{x}}+\cos[\Theta_j]\hat{\mathbf{z}}\,.
    \end{equation}

The resistance\cite{Dugaev2003,Dugaev2006} and STT can now be found by directly solving Schr\"odinger's equation either analytically, where possible, or numerically as follows. Using the general ansatz
    \begin{equation}
    |\psi\rangle=\sum_{j\sigma}\phi_{j\sigma}\psi_{j\sigma}^\dagger|0\rangle\,,
    \end{equation}
where $|0\rangle$ is the vacuum state, one finds that Schr\"odinger's equation $H|\psi\rangle=\epsilon|\psi\rangle$ results in
    \begin{equation}
    -t\left(\phi_{j+1\sigma}+\phi_{j-1\sigma}\right)-\varepsilon_\sigma\phi_{j\sigma}
    -\frac{J}{2}\sum_{\sigma'}\vec\sigma_{\sigma\sigma'}\cdot\vec n_j\phi_{j\sigma'}=0\,,
    \end{equation}
$\forall j,\sigma$, where $\varepsilon_\sigma=\epsilon+\mu$ with $\epsilon$ the energy. For lattice sites in the regions of homogeneous magnetization this is solved by plane wave solutions $\phi_{j\sigma}=\e^{\im k_\sigma ja}$ with eigenenergies
    \begin{equation}
    \varepsilon_\sigma=-2t\cos[k_\sigma a]+ \sigma \frac{J}{2}\,.
    \end{equation}
so that the velocities are 
    \begin{equation}
    u_\sigma=2ta\sin[ k_\sigma a]\,,
    \end{equation}
where $a$ is the lattice spacing. For the scattering problem with a polarized incoming electron from the left we have to solve for the coefficients inside the DW
    \begin{equation}\label{ansatz}
    \phi_{j\sigma}=
    \begin{cases}
          a_\sigma\e^{\im k_\sigma ja}+r_\sigma\e^{-\im k_\sigma ja} & \text{if}\ j<-\ell\,, \\
          \phi_{j\sigma} & \text{if}\ -\ell\leq j\leq \ell\,,\text{ and} \\
          t_{\sigma}\e^{\im k_{\bar\sigma} ja} & \text{if}\ j>\ell\,.
        \end{cases}
    \end{equation}
where $\dw=(2\ell+1)a$ is the width of the DW and $\sigma=-\bar\sigma=\pm1$. In the limit of a sharp DW, $\ell=0$ or equivalently $\dw=a$
    \begin{equation}
    \hat n_j=
    \begin{cases}
          \hat{\mathbf{z}} & \text{if}\ j\leq-1\,, \\
          \hat n_0 & \text{if}\ j=0\,,\text{ and} \\
          -\hat{\mathbf{z}} & \text{if}\ j\geq1\,,
        \end{cases}
    \end{equation}
a fully analytical solution can be found for an arbitrary direction $\hat n_0$
    \begin{eqnarray}\label{transmission}
        t_\sigma&=&\frac{a_\sigma(J+2\sigma \varepsilon_\sigma)(J(1+n^z_0)-2\sigma\varepsilon_\sigma-\im \delta u)}{2J(J-\im\delta u)(1+n^z_0)-\delta u^2-4\varepsilon_\sigma^2}\\\nonumber&&
        +\sigma\frac{a_{\bar\sigma} J (J-2\sigma\varepsilon_\sigma)(n^x_0-\sigma\im n^y_0)}{2J(J-\im\delta u)(1+n^z_0)-\delta u^2-4\varepsilon_\sigma^2}\,,
    \end{eqnarray}
for the transmitted part of the wavefunction where $\delta u=u_1-u_{-1}$.  The tunneling amplitude is then given by 
    \begin{equation}
    	T_\sigma\equiv
    	\frac{ |t_{\sigma} |^2u_{\sigma}}{|a_{1} |^2u_{1}+|a_{-1} |^2 u_{-1} }\,.
    \end{equation}
    
    \begin{figure}[t!]
    \centering
    \includegraphics[width=0.99\columnwidth]{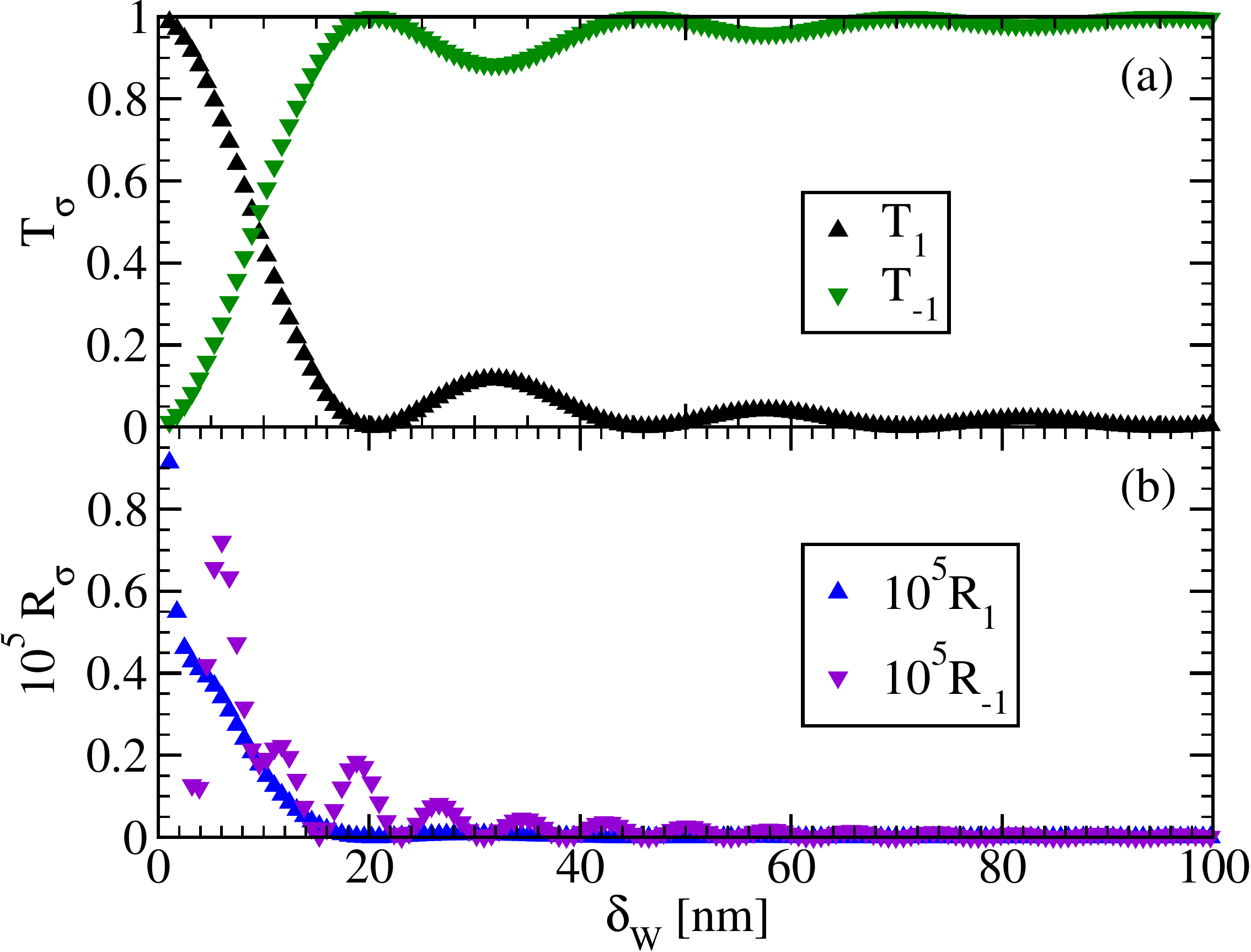}
    \caption{(a) The transmission amplitude $T_\sigma$ and (b) the reflection amplitude $R_\sigma$ for a polarized electron $a_1=1$ at the Fermi momentum as a function of DW width $\dw$.}
    \label{fig:tr}
    \end{figure}

For the more general case of N\'eel DWs with an arbitrary width we use a linear change of the direction
    \begin{equation}
    \Theta_j=
    \begin{cases}
          0 & \text{if}\ j<-\ell\,, \\
          \frac{\pi}{2}\frac{|j+\ell+1|}{\ell+1} & \text{if}\ -\ell\geq j\geq\ell\,,\text{ and} \\
          \pi & \text{if}\ j>\ell\,,
        \end{cases}
    \end{equation}
which can be solved numerically.  Of course any form of the DW is possible here, but the dependence on width $\dw=(2\ell+1)a$ remains the same, while the detailed shape in the plane has little effect on the results. The scattering calculations can then be used to find spin densities, magnetoresistance, and spin transfer torques as described below.

For Permalloy we take $J=0.29$\,eV and $\mu=0.45$\,eV along with $v_{\mathrm{f}}=0.91 \times 10^6$\,ms${^{-1}}$ and $k_{\mathrm{F}}=.159/$\AA.\cite{Reeve2019} The lattice parameter of bulk Ni$_3$Fe is $a=3.55$\,{\AA} and we find
    \begin{equation}
    t= \frac{ \hbar v_{\mathrm{f}} }{ 2 a \sin \left[ k_{\mathrm{F}} a \right] }
    \end{equation}
for the hopping integral of our lattice model. This results in $t=1.58$\,eV and wavevectors $k_1a=1.76$ and $k_{-1}a=1.67$. We use these values through the rest of this article, unless explicitly mentioned otherwise. In Fig.~\ref{fig:tr} we show the transmission and reflection for a polarized incoming particle $a_1=1$ at the Fermi momenta as a function of DW width $\dw$. For the shortest DWs we observe a small amount of reflection and large transmission with the same spin, i.e.~the electrons retain the original polarization without relaxation if the magnetization changes abruptly, which is natural as the electron cannot follow the magnetization quickly. The reflection $R_1$ drops with increasing DW width, but so does the transmission $T_1$ into the same spin channel. Instead the transmission $T_{-1}$ {\it and} reflection $R_{-1}$ into the opposite spin channel increase and dominate for $\dw \agt 18$\,nm. For even longer DW width we enter a second regime, where periodic modulations can be observed, which can be understood as interference effects as the relevant length scales, $\lambda_\sigma\equiv 2\pi/k_\sigma$ and $\lambda_{\pm}=2\pi/(k_1\pm k_{-1})$, become commensurate and incommensurate with the DW width.  In particular, we observe almost perfect transmission $T_{-1}\to 1$ approximately every 15 lattice spacings, which is related to the well known phenomenon of resonant tunneling,\cite{Razavy2003} albeit into the {\it opposite} spin channel.  This implies that the spin polarization of the incoming spin
is coherently and completely transferred to the DW without relaxation or production of heat.  This is far from obvious since naively one could have expected a precession in an arbitrary non-equilibrium state outside the DW, if relaxation is suppressed.
    
    \begin{figure}
    \centering
    \includegraphics[width=0.99\columnwidth]{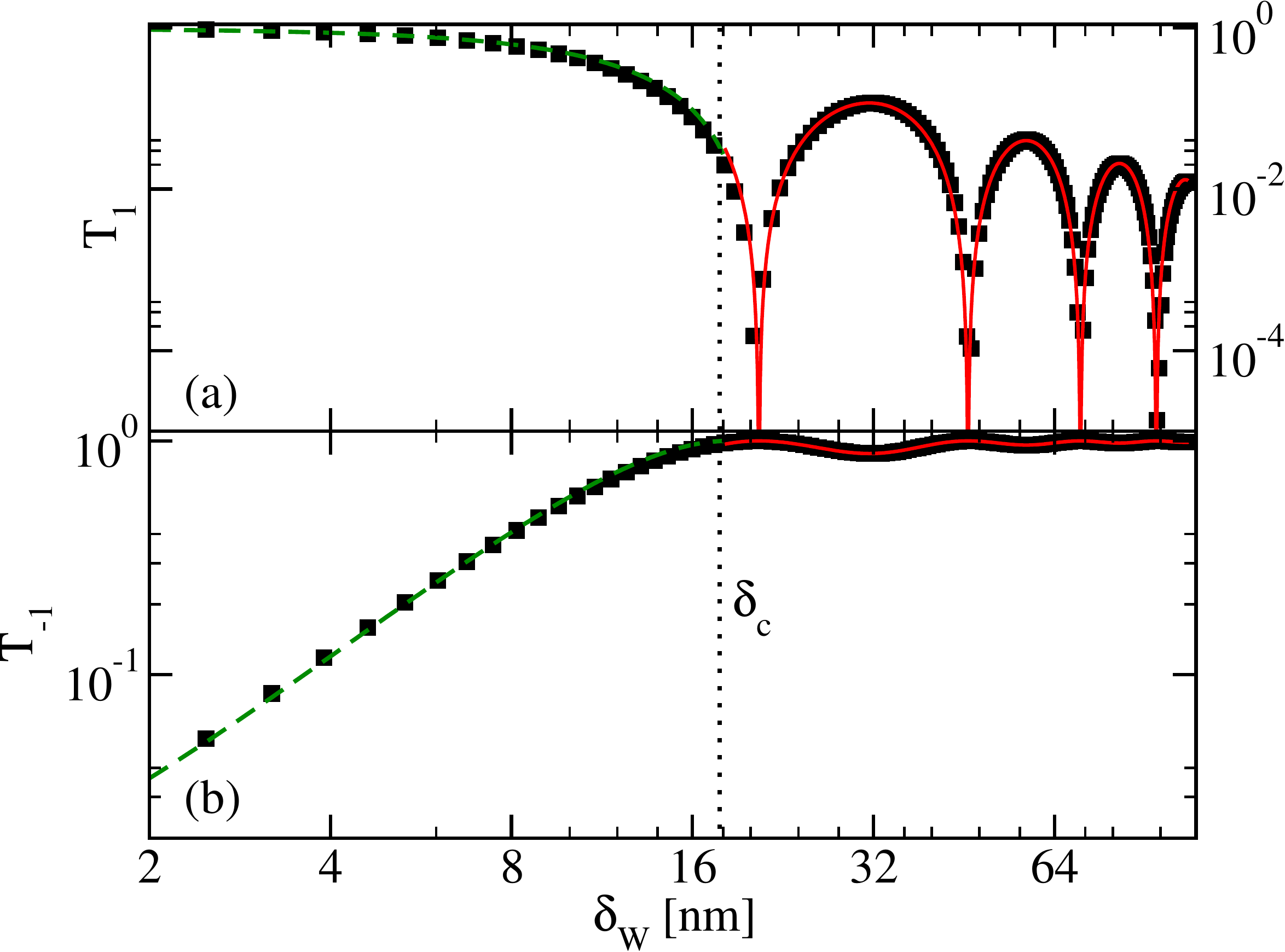}
    \caption{(a) The spin preserving transmission amplitude $T_1$ and (b) the spin flip transmission amplitude $T_{-1}$. Exact results from the numerical and analytical calculations (black squares) are compared to the fits to Eqs.~(\ref{Eq:fit1})-(\ref{Eq:fit}) for $\dw<\dwc$ (dashed green line) and $\dw>\dwc$ (solid red line). The two regimes are clearly visible. See Table \ref{Table:fitValue} for fitting parameters.}
    \label{fig:t}
    \end{figure}

    \begin{figure}
    \centering
    \includegraphics[width=0.99\columnwidth]{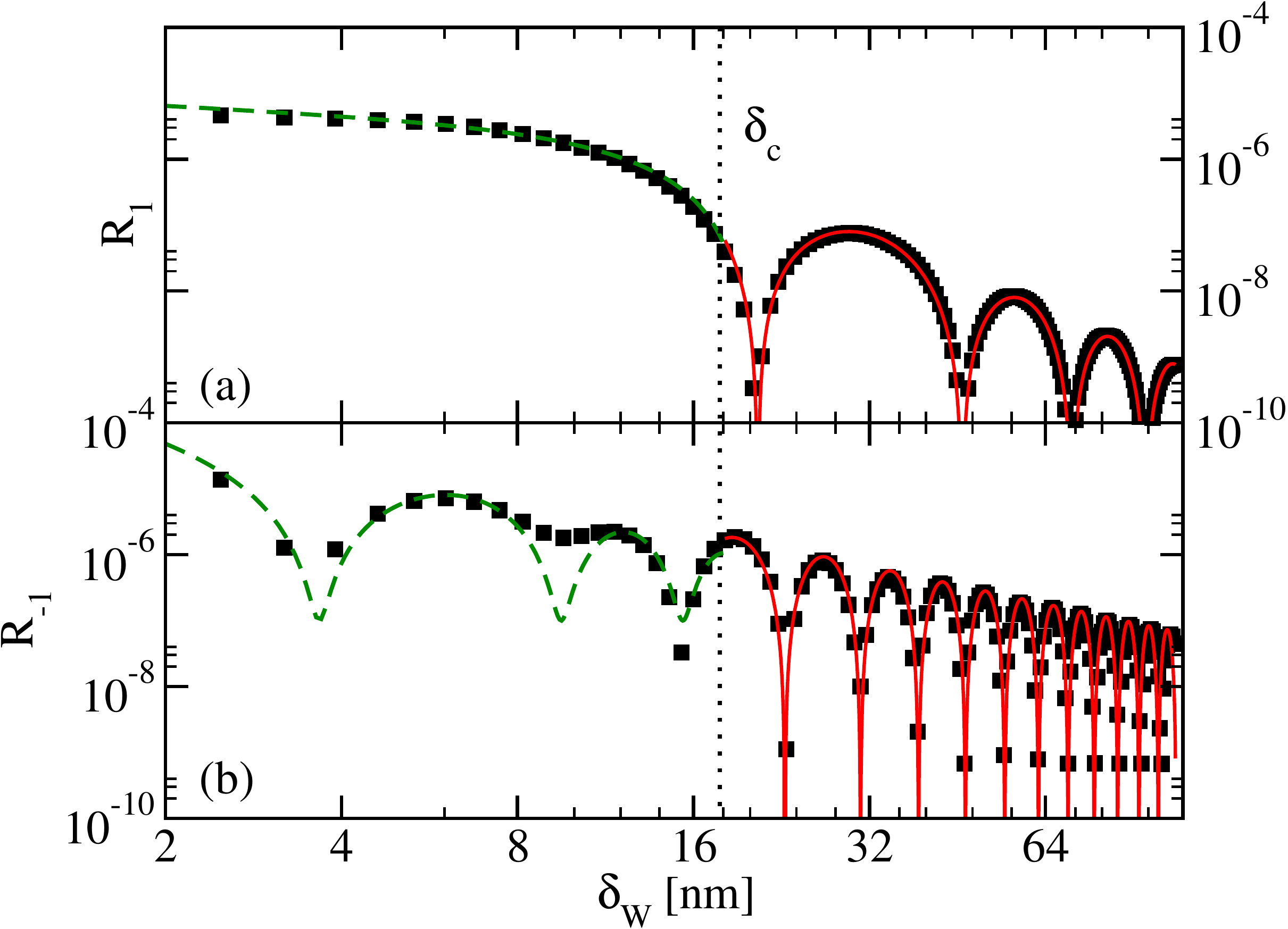}
    \caption{(a) The spin preserving reflection amplitude $R_1$ and (b) the spin flip reflection amplitude $R_{-1}$. Exact results from the numerical and analytical calculations (black squares) are compared to the fits to Eqs.~(\ref{Eq:fit1})-(\ref{Eq:fit}) for $\dw<\dwc$ (dashed green line) and $\dw>\dwc$ (solid red line). See Table \ref{Table:fitValue} for fitting parameters.}
    \label{fig:r}
    \end{figure}
    
A more quantitative analysis is possible in terms of phenomenological fitting functions
describing decaying periodic modulations. As shown in Figs.~\ref{fig:t} and \ref{fig:r} there are two regimes with distinctly different behavior. For shorter DWs $\dw \alt \delta_c =50a=17.75$\,nm we find that all coefficients can be fitted to the form 
    \begin{equation}
    F(\dw\leq \dwc)=
    C+A\left(\frac{\dw}{a}\right)^{-\Gamma}\cos{\left[
    \Omega_0+\Omega_1\frac{\dw}{a}
    \right]}^2. \label{Eq:fit1} 
    \end{equation}
In the short DW limit the abrupt change in magnetization does not allow significant 
transmission into the opposite spin channel, so that the transmission $T_1$ remains large, 
albeit with substantial backscattering with either spin. For longer DWs we observe that $T_{-1}$ dominates, but the misalignment of the electron spin remains large and contributes to the change of the electron motion, so the coefficients are described by a different set of parameters for $\dw > \delta_c$
    \begin{eqnarray}
    F(\dw>\dwc) &=&
    C+B\left(\frac{\dw}{a}\right)^{-\gamma} \nonumber\\&&\times\cos{\left[
    \omega_0 + \omega_1\frac{\dw}{a} +\omega_2\frac{\dw^3}{a^3}
    \right]^2}\,. 
    \label{Eq:fit}
    \end{eqnarray}
We also observe a slight dependence of wavelength on the DW width, so a higher order correction to $\omega_1$ is included here. The values of the fitting parameters $\{c,A,B,\Omega_{0,1},\omega_{0,1,2}\}$ are shown in Table \ref{Table:fitValue}, where the resulting domain wall magneto-resistance (DWMR) will be discussed in the next section.

    \begin{table}
        \begin{tabular}{|p{1.6cm}|p{1cm}|p{1.6cm}|p{1cm}|p{1cm}|p{1.0cm}|}
        \hline
        $F$    &$C$  &$A$   &$\Gamma$ &$\Omega_0$   &$\Omega_1$  \\
           \hline
        $T_1$  &$0$    &$1$  &$0.02$ &$0$   &$0.0276$   \\
        $T_{-1}$    &$1$    &$-0.993$  &$0$ &$0$   &$0.0305$  \\
        $R_1$    &$0$  &$15\cdot10^{-6}$   &$0.49$ &$2.985$   &$0.031$ \\
        $R_{-1}$   &$0$  &$0.00251$   &$2$ &$0.4$   &$0.191$ \\
        $\textrm{DWMR}$   &$0$  &$1$   &$0.2$ &$0$   &$0.026$
        \\\hline\multicolumn{6}{c}{\phantom{.}}\\\hline
        $F$ &$B$  &$\gamma$ &$\omega_0$   &$\omega_1$   &$\omega_2$   \\
        \hline
        $T_1$ &$494$  &$1.85$ &$2.18$   &$0.435$   &$2\cdot10^{-8}$   \\
        $T_{-1}$  &$494$  &$1.85$ &$2.18$   &$0.435$   &$2\cdot10^{-8}$   \\
        $R_1$&$0$  &$3.8$ &$2.2$   &$0.0432$   &$2\cdot10^{-8}$   \\
        $R_{-1}$   &$0.0033$  &$1.89$ &$2$   &$0.139$   &$6\cdot10^{-8}$   \\
        $\textrm{DWMR}$   &$5720$  &$2.7$ &$.72$   &$0.04$   &$5\cdot10^{-8}$   \\
        \hline
        \end{tabular}
        \caption{Fitting parameters for the scattering amplitudes in Figs.~\ref{fig:t} \ref{fig:r}, and \ref{fig:dwmr_biased}. The critical width is $\dw^{\rm c}=50a$.}
        \label{Table:fitValue}
    \end{table}

In the first regime $\dw<\delta_c$  the exponent $\Gamma$ is very small, i.e.~there is very little drop at small DW widths for the spin flip transmission, since it takes a longer DW for the electrons to change magnetization. In the second region $\dw>\dwc$ the exponent is much larger $\gamma \approx 2$. This value is very close to the value predicted by analytical methods for the behaviour of the DWMR in the adiabatic regime which closely relates to these scattering rates. For $\dw > \delta_c$, all amplitudes decay with a rate $\gamma \approx 2$, except reflection into the same channel which decays approximately twice as fast $\gamma = 3.8$. 

The inverse modulation lengths $\omega_1$ and $\Omega_1$ become larger for particles with higher incident energies as expected as the relevant length scales are shorter. In Appendix \ref{app:fit} we show how the scattering amplitudes change for different incoming electron momenta.

\section{Resistance}\label{sec:resistance}

    
We now consider the resulting resistance through a clean wire connected to two reservoirs on the left and right held at slightly different biases. The resistance for each incident particle can be determined from
    \begin{equation}
    	R=R_1+R_{-1}=
    	\frac{ |r_{1} |^2u_{1}+|r_{-1} |^2u_{-1}}{|a_{1} |^2u_{1}+|a_{-1} |^2 u_{-1} }\,,
    \label{res}
    \end{equation}
with the coefficients from the solutions to the scattering problem.

    \begin{figure}
    \centering
    \includegraphics[width=0.99\columnwidth]{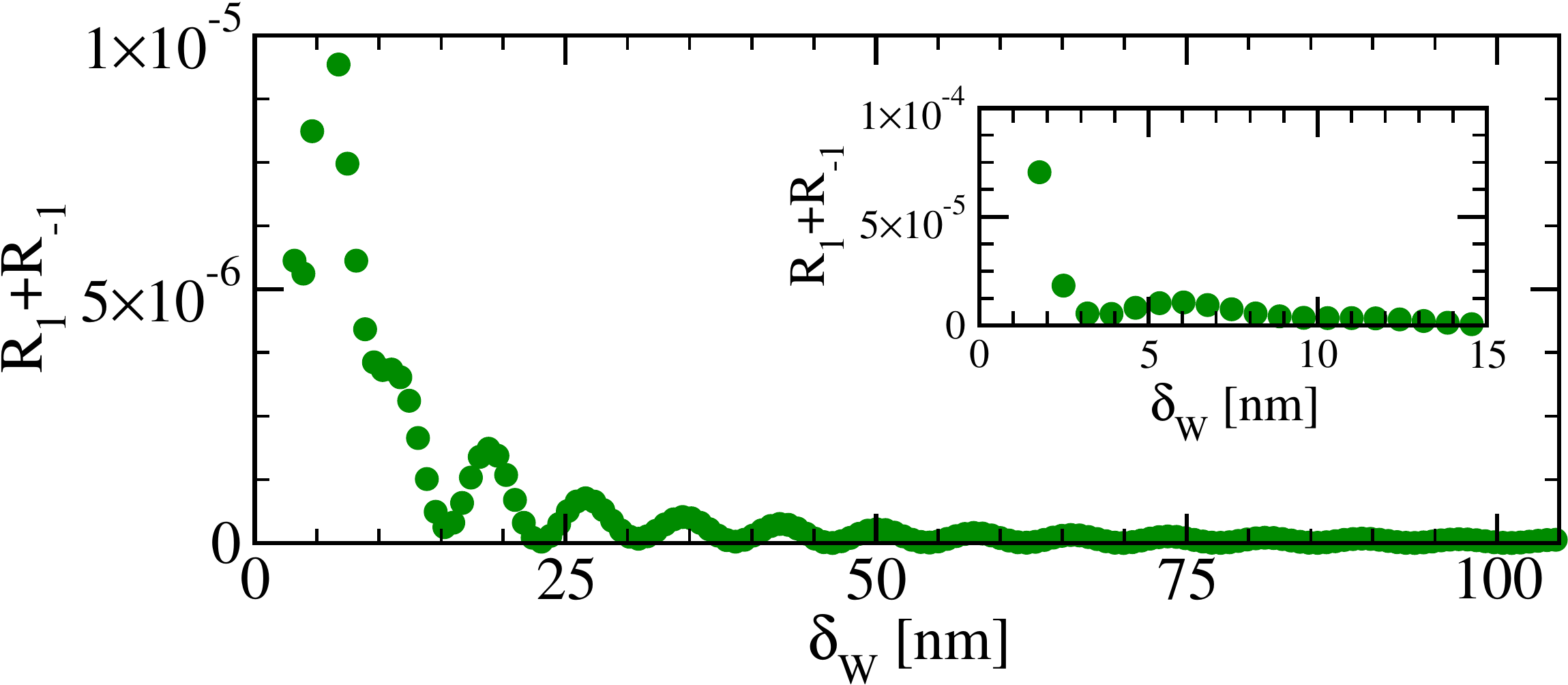}
    \caption{Behavior of the DW resistance from Eq.~(\ref{res}) at the Fermi momenta as a function of DW width $\dw$. For shorter DWs (see inset) the modulations are sizable and for longer DW widths backscattering becomes very small.}
    \label{fig:dwmr_vs_dwl}
    \end{figure}

    \begin{figure}
    \centering
    \includegraphics[width=0.99\columnwidth]{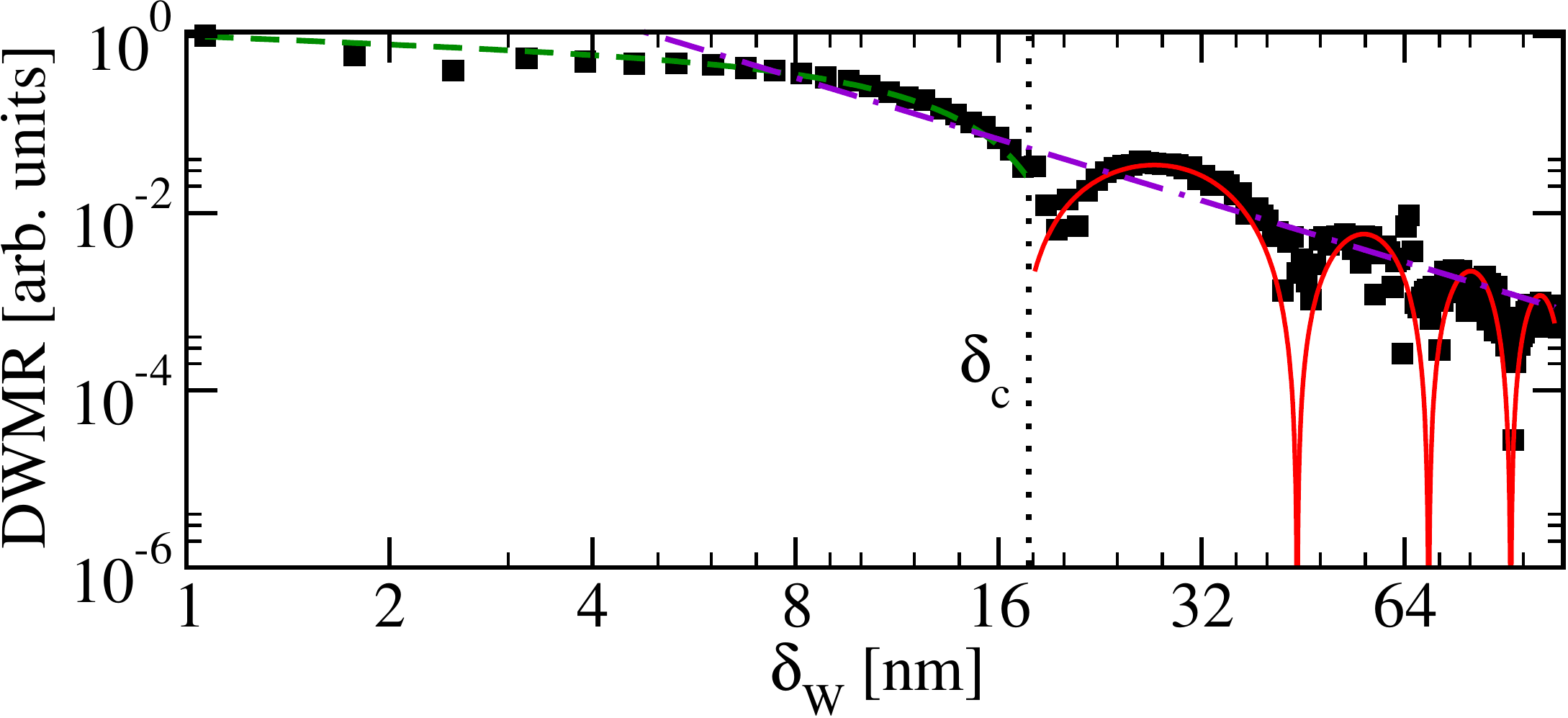}
    \caption{DWMR as a function of DW width $\dw$, and several fits to the numerical data. The black squares visualize the numerical data which are normalized by the maximum value of the obtained DWMR. The magenta line is the fit for $\dw<\dw^\mathrm{c}$ and the cyan line is the fit for $\dw>\dw^\mathrm{c}$ to Eqs.~(\ref{Eq:fit1})-(\ref{Eq:fit}). The blue line is the fitted line given by the function $C\dw^{-\Bar{\gamma}}$ with the experimental value of the decay-exponent $\Bar{\gamma}=2.3$. The blue line fits very well to the maxima of the DWMR over the sample. Effective Permalloy parameters were used. See Table \ref{Table:fitValue} for fitting parameters.}
    \label{fig:dwmr_biased}
    \end{figure}

In order to find the total DWMR we consider particles at the Fermi-energy but with arbitrary incident direction on the DW, with the momentum perpendicular to the DW conserved. Accordingly, the DWMR is given by the average over the resistance $R(\vec{k}_\sigma)$ for different incident particles with momentum $\vec{k}_\sigma$ over the 3D Fermi sphere~\cite{Ashcroft1976,Stiles2002},
    \begin{equation}\label{eq:theoDWMR}
    	\frac{1}{R_{\rm DW}}=e^2 \int \left[ \frac{d \vec{k}_\sigma}{8\pi} \frac{1}{R(k_\sigma)} \nu(\vec{k}_\sigma)\left( \frac{\partial f(\vec{k}_\sigma)}{\partial \vec{k}_\sigma} \right) \right],
    \end{equation}
where $e$ is the electron charge, $\nu(\vec{k})$ is the velocity, and $f(\vec{k})$ is the Fermi-Dirac distribution. In our case, the scattering parameters only depend on the $\vec{k}_\sigma$-components in the direction of changing magnetization if the perpendicular components are approximately conserved. It is therefore sufficient to use the 1D scattering result in Eq.~(\ref{res}) from the
plane-wave ansatz in Sec.~\ref{sec:nanocontacts}. We will consider here an unpolarized spin-current, $a_\sigma=\frac{1}{\sqrt{2}}$.

From Eq.~(\ref{eq:theoDWMR}) and after going to polar coordinates, the resistance for each $\dw$ averaged over different momenta is simplified as
    \begin{equation}
        \frac{1}{R_{\rm DW}} = \int_0^1 {\rm d}z k_{\rm F}^2
        \frac{v(k_{\rm F} z)^2}{R(k_{\rm F} z)}\,,
    \label{eq:totalDWMR} 
    \end{equation}
where $z=\cos{\theta}$ parameterizes the incoming direction and we have approximated $\frac{\partial f(\vec{k}_\sigma)}{\partial \vec{k}_\sigma}$ with a delta function which is justified for the relevant parameter ranges. Since this is solved for a number of incident particles with different momenta numerically, the integral is turned into a summation over the values obtained for a reasonably large number of incident particles (here we found out that 150 different $k$-values suffice). 

In Fig.~\ref{fig:dwmr_vs_dwl} we show the dependence of the DW resistance on $\dw$. Large periodic modulations with the DW width can be seen, which persist up to long DWs. These modulations are still apparent in the DWMR after averaging over all particle-directions in Fig.~\ref{fig:dwmr_biased}.  Fits to decaying modulations in the different regimes are also shown, which yield the values in Table \ref{Table:fitValue}, where the exponent is now slightly larger than 2.
    
\section{Spin Density}\label{sec:spindensity}

\begin{figure*}
    \centering
    \includegraphics[width=1.79\columnwidth]{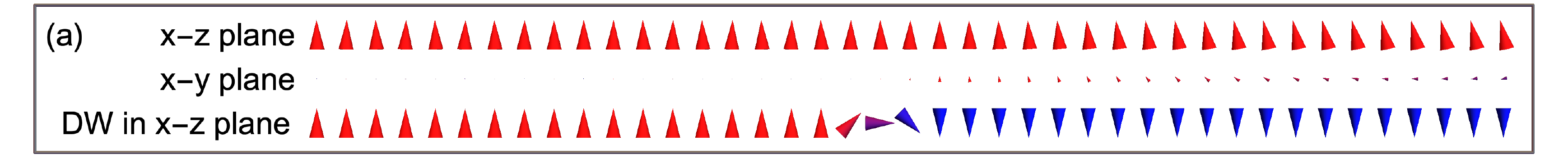}\\
    \includegraphics[width=1.79\columnwidth]{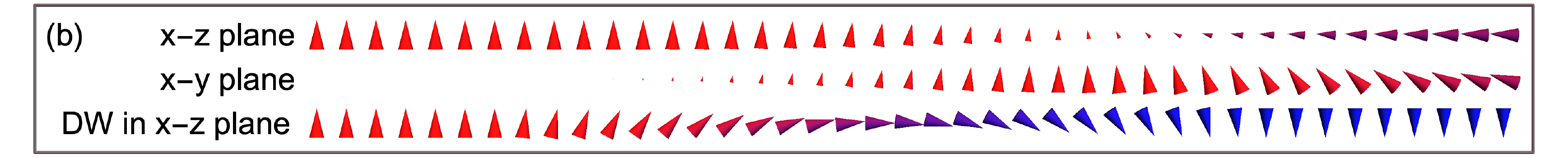}\\
    \includegraphics[width=1.79\columnwidth]{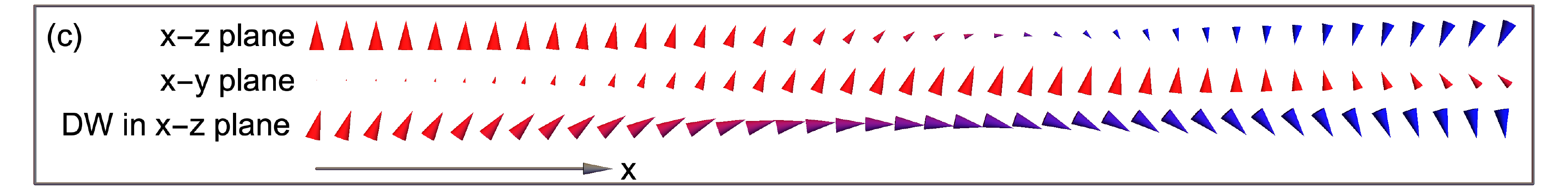}
    \caption{The orientation of three N\'eel domain walls confined to the $x$-$z$ plane and the conduction electrons' spin density in the $x$-$z$ plane and in the $y$-$z$ plane. Even for the longer DWs the conduction electrons spin is not collinear with the DW. The DW widths are: (a) $\ell=1$ ($\delta_{\textrm{DW}}=3a$), (b) $\ell=12$ ($\delta_{\textrm{DW}}=25a$), and (c) $\ell=22$ ($\delta_{\textrm{DW}}=45a$).}
    \label{fig:dw_vs_spin_density}
\end{figure*}

To get a deeper insight into the time-evolution and to predict the STT, we now consider the total spin density at site $j$ 
    \begin{equation}
    {\delta \vec S}_{j\tau}=\frac{1}{2}\sum_{\sigma\sigma'}\phi^*_{j\sigma}\vec\sigma_{\sigma\sigma'}\phi_{j\sigma'}|_{a_\tau=1,a_{\bar\tau}=0}\,,
    \end{equation}
where $\tau=\pm1= -\bar \tau$ specifies the polarization of the incoming current. For the sharp DW limit $\dw=a$ we can again find analytical expressions, including at the site of the DW $j=0$
    \begin{eqnarray}
    \delta S^x_{\tau}&=&\tau\frac{J(J+2\tau\varepsilon)^2\left[n^x_0(J-2\tau\varepsilon)-\tau\delta u n^y_0\right]}{2\left[(\delta u^2+4\varepsilon^2)^2+4J^4-16J^2\varepsilon^2\right]}\,,\\\nonumber
    \delta S^y_{\tau}&=&\tau\frac{J(J+2\tau\varepsilon)^2\left[n^y_0(J-2\tau\varepsilon)+\tau\delta u n^x_0\right]}{2\left[(\delta u^2+4\varepsilon^2)^2+4J^4-16J^2\varepsilon^2\right]}\,,\textrm{ and}\\
    \delta S^z_{\tau}&=&\tau\frac{(J+2\tau\varepsilon)^2\left[\delta u^2+4\varepsilon^2-4\tau J\varepsilon\right]}{2\left[(\delta u^2+4\varepsilon^2)^2+4J^4-16J^2\varepsilon^2)\right]}\,.\nonumber
\end{eqnarray}
Already in this limit it becomes clear, that a N\'eel DW confined to the $x$-$z$ plane, nonetheless produces spin densities in all directions. In fact for the ratio
    \begin{equation}
        \Gamma_\tau\equiv\left|\frac{\delta S^y_\tau
        }{\delta S^x_\tau}\right|=\left|\frac{\sin k_1-\sin k_{-1}}{2\cos k_{\bar\tau}}\right|
    \end{equation}
one finds $\Gamma_1=0.0344$ and $\Gamma_{-1}=0.0653$. Thus we expect spin torques which tend to rotate the DW out-of-plane as well as terms which tend to move it along the plane. As discussed in the next section both torques are relevant even without relaxation processes.

In Fig.~\ref{fig:dw_vs_spin_density} we show examples of the spin density of the conduction electrons compared to the DW orientation for several exemplary DW lengths. In all cases the spin density of the electrons can be seen to deviate far from collinearity with the DW orientation. This is true even for the case $\ell=22$ in which the DW width is $45a$, much longer than the relevant wavelengths $\lambda_1=3.57a$ and $\lambda_{-1}=3.78a$. We note here the other relevant length scales for scattering are $\lambda_+=1.84a$ and $\lambda_-=62.8a$.

Similar to the sharp DW case the fact that the spin density is not close to collinearity with the DW ensures that there will be a complicated local spin torque acting on the DW. Under the assumption that the DW dynamics are very slow on the time scale of electron motion, the DW orientation can be taken as quasi-stationary allowing one to solve the equilibrium scattering problem for a stationary DW profile to find the spin densities.

\section{Spin transfer torques}\label{sec:STT}

In the $s$-$d$-model the spin torque acting on the DW can be deduced from the spin lost by the conduction electrons. The STT can therefore be found from the Heisenberg equation for the spin of the conduction electrons: $\tfrac{\dD \boldsymbol{S}_i}{\dD t}=\im [\hh,\boldsymbol{S}_i]$, where after quantum mechanical averaging, $\langle\ldots\rangle$, we have
    \begin{equation}\label{contieq}
        \frac{\partial \langle \boldsymbol{S}_i \rangle}{\partial t} = 
        \langle \boldsymbol{J}^{\rm s}_i \rangle - \langle \boldsymbol{J}^{\rm s}_{i-1}\rangle + J \boldsymbol{n}_i \times \langle\boldsymbol{S}_i\rangle\,.
    \end{equation}
Here the spin current $\boldsymbol{\hat{J}}^{\rm s}$ is defined as
    \begin{equation}
        \boldsymbol{J}^{\rm s}_i = 
        -\im t \sum_{\sigma \sigma'} 
        \left(  
        c^{\dagger}_{i+1\sigma} \boldsymbol{S}_{\sigma\sigma'} c_{i\sigma'} - c^{\dagger}_{i\sigma} \boldsymbol{S}_{\sigma\sigma'}c_{i+1\sigma'}
        \right)\,. 
    \end{equation}
In the steady state, the time derivative vanishes and the change of spin-current must be
entirely due to the torque, which in turn is therefore given by the last term in Eq.~(\ref{contieq})
	\begin{equation}\label{eq:stt}
	    \boldsymbol{T}_i=-J\boldsymbol{n}_i \times\boldsymbol{S}_i\,,
	\end{equation}
where we choose to omit relaxation in order to identify all torques from the quantum equation of motion. Note that the torque is perpendicular to the magnetization, but can be either in the $x$-$z$ plane or perpendicular to it, depending on the misalignment of the electron spins. For an almost uniform magnetization it is commonly argued that the in-plane component is due to the adiabatic change of motion,\cite{Stiles2002} while the out-of-plane torque comes from 
non-adiabatic relaxation processes.\cite{Zhang2004}  This is no longer true for a changing magnetization, where we observe in-plane and out-of-plane torques both for a very abrupt change but also for longer DWs, which can be traced to an effective gauge field in the 
frame of reference of the magnetization.\cite{Bazaliy1998}

\begin{figure}
    \centering
    \includegraphics[width=0.99\columnwidth]{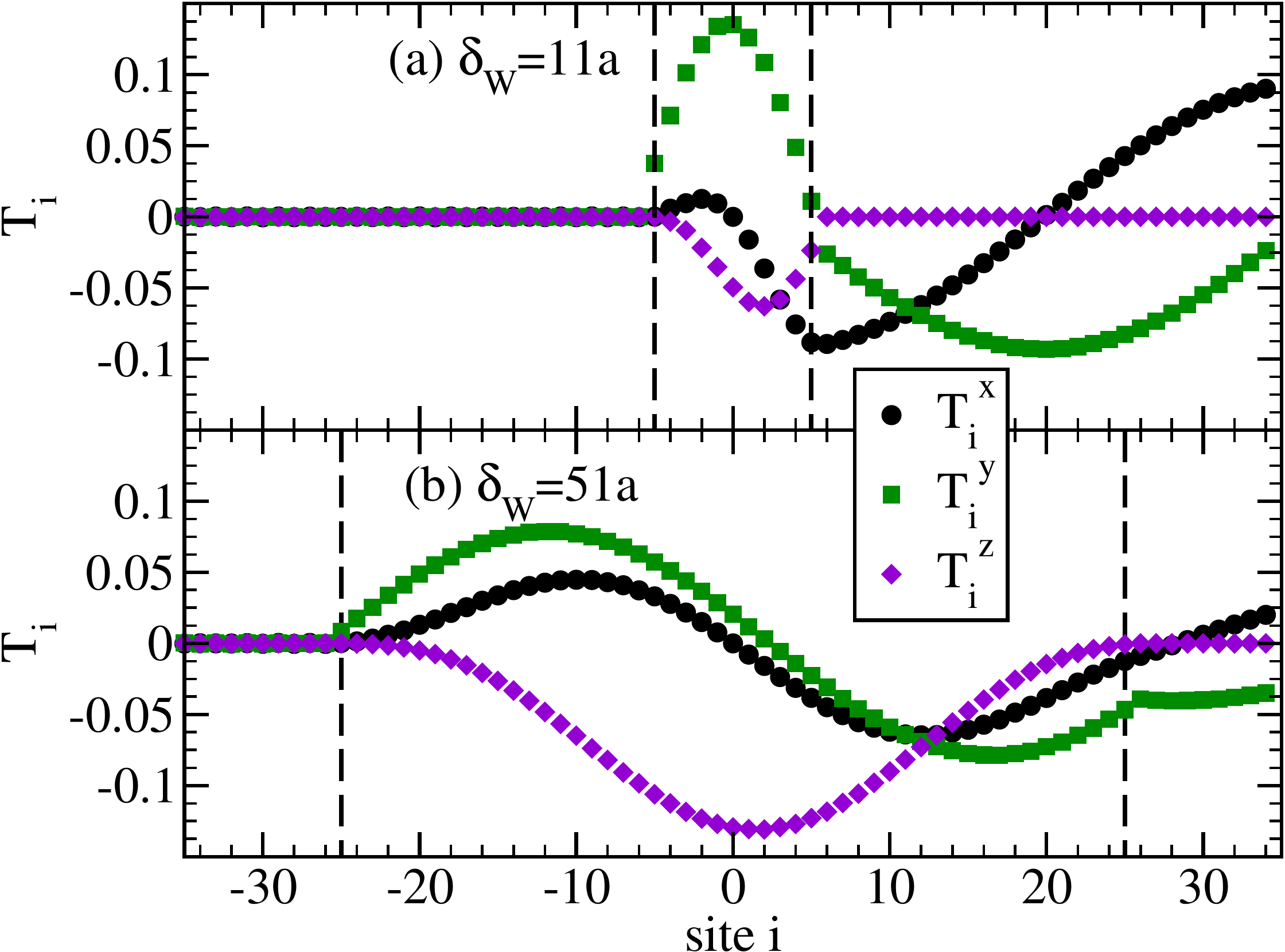}
    \caption{The STT caused by the misalignment of the spin density with the DW profile per site, Eq.~\eqref{eq:stt}, for DWs of width (a) $\dw=11a$ and (b) $\dw=51a$. The limits of the DW are shown by dashed lines.}
    \label{fig:stt_sites}
\end{figure}

In Fig.~\ref{fig:stt_sites} we show this torque over the whole DW region for two exemplary DW widths. As reflection of the electrons is small and the incident electrons are polarized parallel to the bulk magnetization direction, there is little torque on the left hand side. However the transmitted electrons may apply a torque to the bulk magnetization of the right of the DW. Over the region of the DW a large torque is produced which has components in all three directions of a similar magnitude. The torque in the $y$ direction would distort the DW out of plane, and corresponds to ordinary precessional motion for an in-plane misaligned electron spin, as in the commonly assumed scenario for uniform magnetization.  However, the in-plane forces are just as large and tend to move the DW. The spatial dependence across the DW width of all terms will start to deform the DW for longer times. It has been seen elsewhere that such spatially varying torques can lower the threshold for lateral DW motion.\cite{Sedlmayr2011c}
    
The $y$-component of the STT is nonzero for short DWs due to mistracking between the DW orientation and the electron spin density.  In longer, slowly varying DWs where mistracking should be small this contribution remains sizeable, which can be associated with the slow change of the magnetization direction and the geometrical phase associated with it.\cite{Bazaliy1998,Xiao2006}

\begin{figure}
    \centering
    \includegraphics[width=0.99\columnwidth]{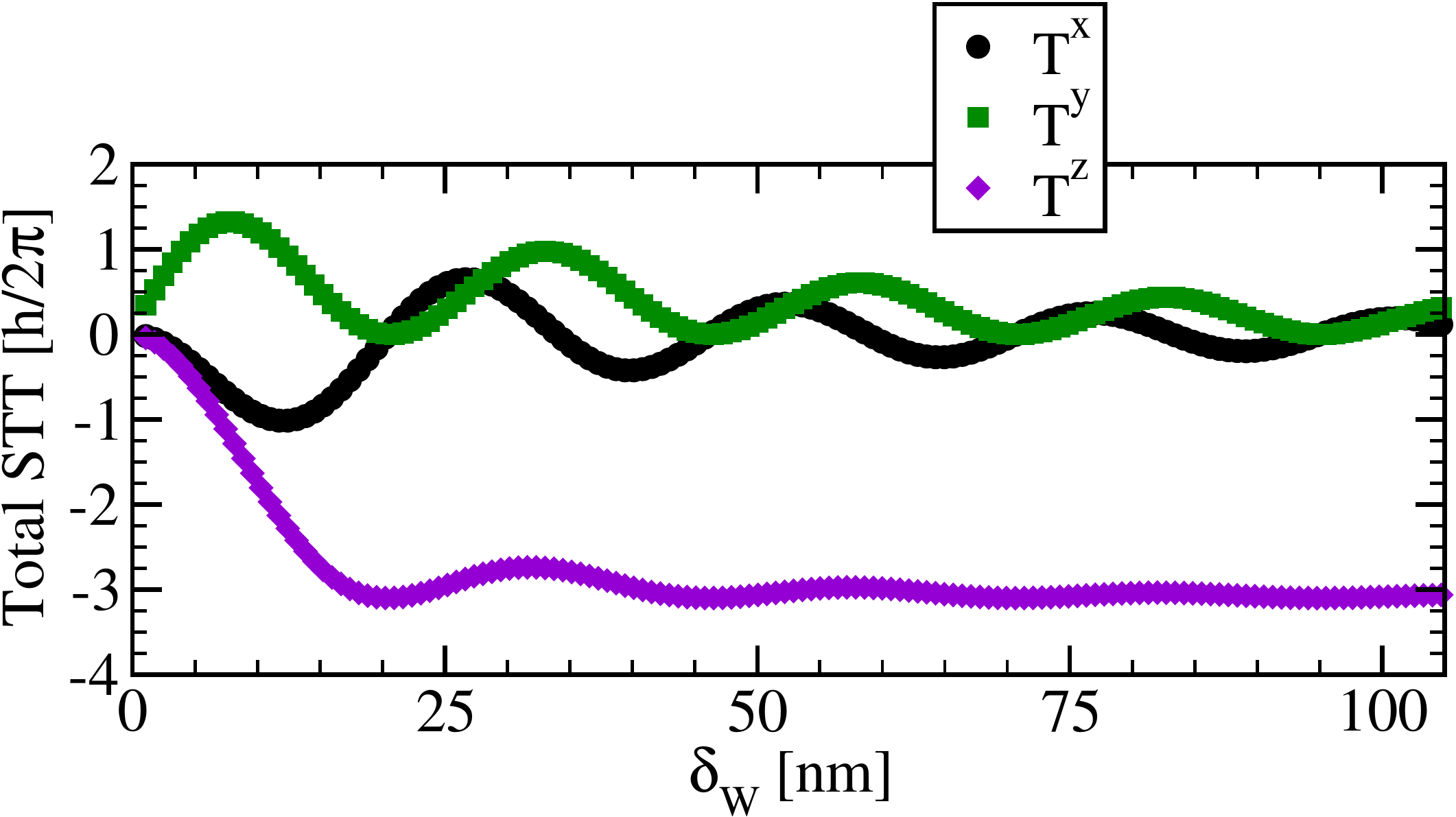}
    \caption{The total STT applied to a DW from spin density misalignment as a function of the DW width.}
    \label{fig:stt_total}
\end{figure}

The total STT applied to the DW is obtained by summing over the DW width as shown in Fig.~\ref{fig:stt_total} as a function of the DW width $\dw$. Because spin is conserved
this is closely related to the transmission coefficient $T_{-1}$, so that in the long DW limit and for resonant widths $T_{-1} \to 1$ the total transferred spin always moves the DW without the need of relaxation or production of heat.   However, in addition the relatively large variation of strong forces will lead to relevant local distortions.

\section{Conclusions}\label{sec:conclusions}

We have investigated the DWMR and the STT in systems with changing magnetization using microscopic simulations of electron transport. To illustrate the effects we use the particular example of 
DWs of variable width $\dw$ in Permalloy. Modulations with DW width $\dw$ are observed in the transport coefficients, which can be traced to commensurate and incommensurate behavior of the relevant length scales. For very short DWs the electron spin is not transferred along the magnetization direction. Accordingly, two different physical regions for short DW and long DW are identified, which are characterized by distinct modulation lengths and decay with DW width. By averaging over all incoming electron directions the complete DWMR can be calculated, which also exhibits periodic modulations with DW width. These results could be used for fine tuning the DWMR, especially if shorter DWs or geometrical pinning are employed. 

In a wide range of DW widths we find that the adiabatic approximation misses important effects from the spin density of the conduction electrons, which is no longer even approximately collinear with the bulk magnetization.  This is reflected in the corresponding local STT, which were analyzed as a function of position along the DW.  Both in-plane and out-of-plane torques are significant 
without assuming any relaxation mechanisms and the complete spin of polarized incoming
electrons can be transferred to the DW by coherent time evolution even for a short DW width
$\dw \sim 20$\,nm.  This resonance effect is notable since it implies that the spin-polarization of the incoming spin can be transferred coherently to the DW without 
relaxation or heat production.  Equally interesting are the large oscillations of strong local torques in 
all directions in both short and long DWs, which will lead to significant DW distortions and movements on longer time-scales. Understanding the microscopic behavior of the local torques therefore paves the way for engineering efficient magnetization manipulations on the nanoscale.


\acknowledgments

We gratefully acknowledge helpful discussions with Bertrand Dup\'e, Vitalii Dugaev, Mathias Kl\"aui, Robert Reeve, Jairo Sinova, Axel Pelster, and Kevin J\"agering. This work was supported by the NAWA/DAAD bilateral grant PPN/BIL/2018/1/00130, as well as the Deutsche Forschungsgemeinschaft (German Research Foundation), Project No. 268565370/TRR173 through the collaborative research center SFB/TRR 173 Spin+X, project A10).


\appendix

\section{Momentum dependence of scattering amplitudes}\label{app:fit}

\begin{figure}[H]
    \centering
    \includegraphics[width=\columnwidth,height=0.6\columnwidth]{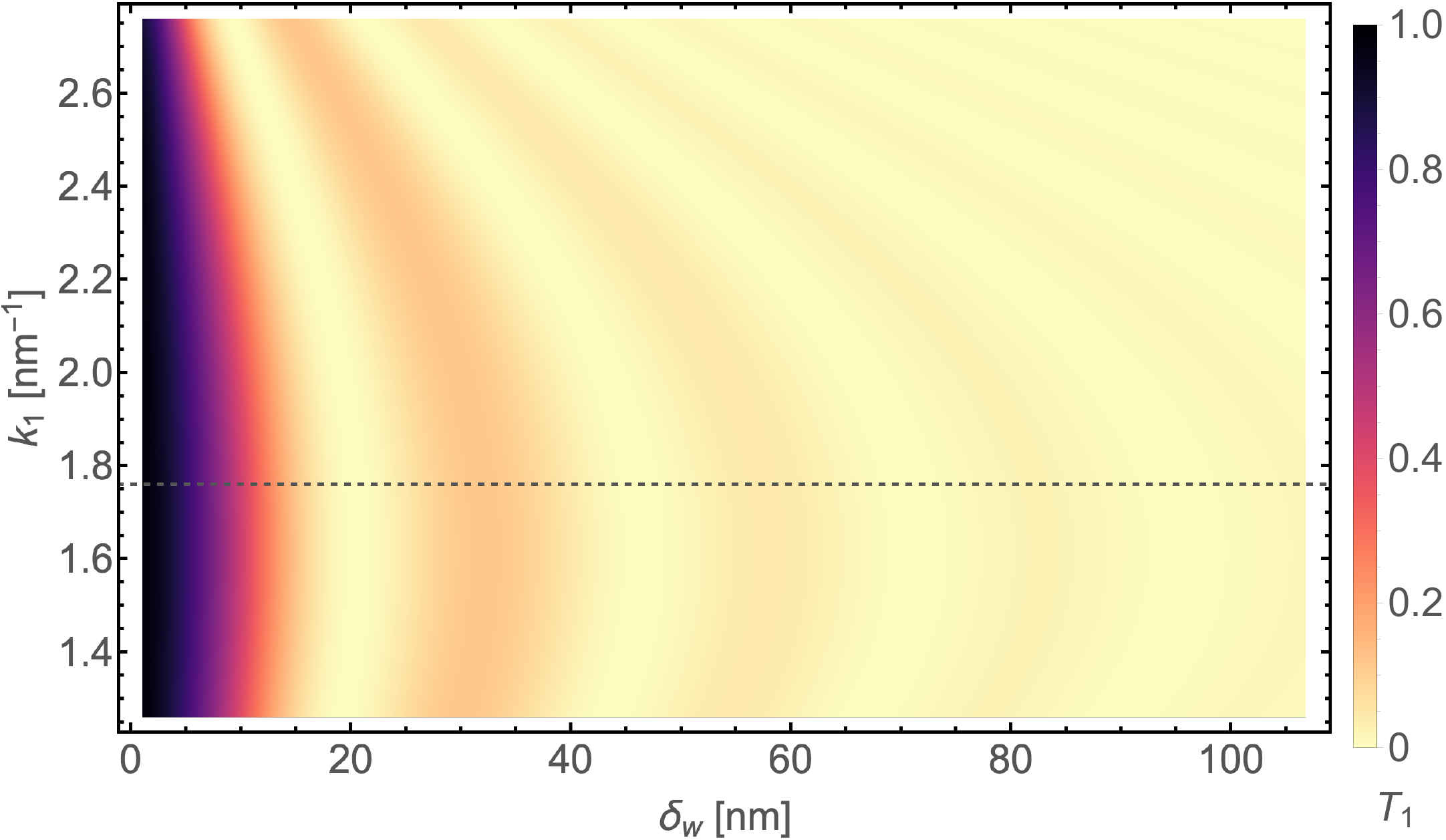}
    \caption{The spin preserving transmission amplitude as a function of the DW width and incoming electron momentum $k_1$. The dashed line shows the momentum used for the Permalloy calculations in the main text.}
    \label{fig:09}
\end{figure}

\begin{figure}[H]
    \centering
    \includegraphics[width=\columnwidth,height=0.6\columnwidth]{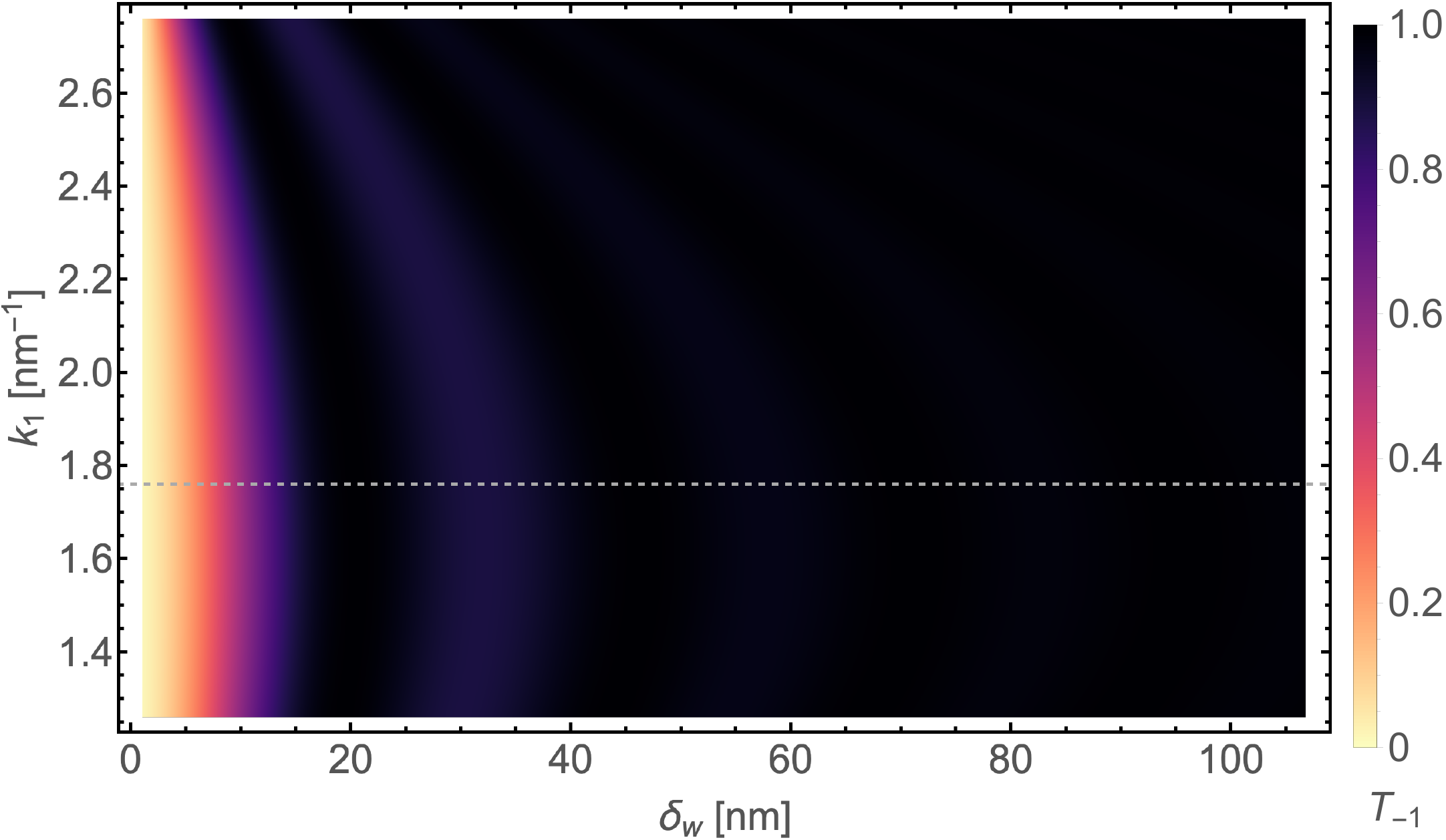}
    \caption{The spin flip transmission amplitude as a function of the DW width and incoming electron momentum $k_1$. The dashed line shows the momentum used for the Permalloy calculations in the main text.}
    \label{fig:10}
\end{figure}

Figs.~\ref{fig:09} and \ref{fig:10} show the momentum and DW width dependence of the scattering amplitudes.

\phantom{.}





\end{document}